\documentstyle[aps,epsf,preprint]{revtex}
\def\a{\alpha}
\def\o{\omega}
\def\g{\gamma}
\def\f{\frac}
\def\p{\phi}
\def\r{\rho}
\def\vl{\vline}
\def\be{\begin{equation}}
\def\ee{\end{equation}}
\def\ba{\begin{eqnarray}}
\def\ea{\end{eqnarray}}
\def\no{\nonumber \\}
\begin{document}
\tightenlines
\begin{titlepage}
\pagestyle{empty}
\vspace{1.0in}
\begin{flushright}
May 1997
\end{flushright}
\vspace{1.0in}
\begin{center}
\begin{large}
{\bf{p-Branes and Exact solutions in Brans-Dicke cosmology with matter}}
\end{large}
\vskip 1.0in
Seung-geun Lee and Sang-Jin Sin\\
\vskip 0.2in
{\small {\it Department of Physics, Hanyang University\\
Seoul, Korea}}
\end{center}
\vspace{1cm}

\begin{abstract}
We study the effect of the solitonic degree of freedom in string cosmology
following the line Rama.
The gas of  solitonic  p-brane is treated as a perfect fluid 
in a Brans-Dicke type theory. 
 In this paper, we find analytic solutions 
for a few solvable  cases of equation of states 
with general Brans-Dicke parameter $\omega$ 
and study the cosmology of the solutions. 
In some cases the solutions are free from initial singularity.
\end{abstract}

\hspace{.2in}PACS numbers:98.80.Cq, 11.25.-w, 04.50.+h
\end{titlepage}

\section{Introduction}
Recent development of the string theory suggest that in a 
regime  of plank length curvature, quantum fluctuation is very large  
so that string coupling becomes large and consequently 
the fundamental string degrees of freedom
are not a weakly coupled {\it good} ones\cite{witten}.
Instead, solitonic degree of freedoms like p-brane or 
D-p-branes\cite{pol} are more important. 
Therefore it is a very interesting question 
to ask what is the effect of these new degrees of freedom to 
the space time structure especially whether including these degree 
of freedom reslove the initial singularity, which is a problem in 
standard general relativity.

For the investigation of  the p-brane cosmology, 
the usual low energy effective action coming from the beta 
function of the string worldsheet would not a good starting point.
So there will be a difference from the string cosmology\cite{vene}. 
We need to  find the low energy effective theory that contains gravity and 
at the same time revealing the effect of these solitonic objects.
But those are not known. Therefore one  can only  guess the answer at this 
moment. 
The minimum amount of the requirement is that it should give a gravity theory 
therefore it must be a generalization of general relativity. The 
Brans-Dicke theory is a generic deformation of the general relativity 
allowing varaible gravity coupling. 
In fact low energy theory of the fundamental string 
gives a Brans-Dicke theory with fine tuned deformation parameter ($\omega$=-1) 
when we look only dilaton field. Moreover, Duff and et.al 
\cite{du} find that, the natural metric that couples to the p-brane is the 
Einstein metric multiplied by certain power of dilaton coupling. 
In terms of this new metric, the action that gives the p-brane solution 
becomes Brans-Dicke action with definite deformation 
parameter $\omega$. 
Using this action, Rama \cite{ra} recently  argued
that the gas of  solitonic  p-brane\cite{du} treated as 
a perfect fluid type matter in a Brans-Dicke theory can resolve the 
initial singularity without any explicit solution. 
In this paper, we find analytic solutions
for some equation of state parameter 
$\gamma$ and for general Brans-Dicke parameter $\omega$ that is 
larger than -3/2 and study cosmology of the solutions. 
Our solution support Ramas claim partially but not in detail.

The rest of this paper is organized as follows.
In section 2,  we set up the action for the p-brane cosmology. 
In section 3, we find an anlytic soltion to the equation of motion and 
constraint equations for a few solvable  cases and discuss the cosmology 
of the solutions.
In section 4, we study the cosmology in Einstein frame for comparison.
In section 5, we summarize and conclude with some discussions.

\section{Effective action with matter}

Let us begin with the effective action, 
\be 
S = \int d^4x {\sqrt {-g} } e^{-\phi}  [ R - \omega   
 \partial_\mu\phi\partial^\mu\phi  - \frac {H^2}{12} ] + S_m ,
\ee 
where $\phi$ is the dilaton field and $H$ is 3-form field strength. 
There can be more fields from 
compactification, higher loop effect, etc. For a moment, we
do not consider other fields
except graviton and dilaton.
Without H-field it is the same as Brans-Dicke(BD) action. It is interesting
to note that in string theory the BD parameter $\omega=-1$. 
In the high curvature regime, the coupling is also big and the
solitonic p-brane will be copiously produced since they become light and
dominates the universe in that regime.
Duff et al \cite{du}
have shown that in terms of metric which couples minimally to 
p-brane$(p=d-1)$, the effective action can be written as Brans-Dicke 
theory the parameter $\o$
\be
\o = -\f{(D-1)(d-2)-d^2}{(D-2)(d-2)-d^2}
\ee
In 4-dim. $\o = -\f{4}{3}$ for particle$(p=0)$ and and $\o=-\f{3}{2}$ for
instanton$(p=-1)$. In 10-dim. $\o = -\f{10}{9}$ for particle and 
$\o =-\f{9}{8}$ for instanton.  
We begin with the effective action with $H=0$. 
\be
S= \int d^4x \sqrt{-g} e^{-\p} (R - \o (\partial\p)^2) + S_m
\ee
By varying $g_{\mu\nu}$ and $\p$, we obtain equations of motion.
\ba
R_{\mu\nu} - \frac{1}{2}g_{\mu\nu}R &=& \frac{1}{2}e^\phi T_{\mu\nu} 
    + \omega ({\partial_{\mu}\phi\partial_{\nu}\phi}
  -\frac{1}{2}g_{\mu\nu}(\partial\phi)^2)  \no
&+& (-\partial_{\mu}\partial_{\nu}\phi +\partial_{\mu}\phi\partial_{\nu}\phi
+g_{\mu\nu}\partial_{\mu}\partial^{\mu}\phi
-g_{\mu\nu}(\partial\phi)^2 ) \\
R - 2\omega\partial_{\mu}\partial^{\mu}\phi &+&
\omega(\partial\phi)^2 = 0 \\
T_{\mu\nu} = p g_{\mu\nu} &+& (p+\rho)U_{\mu}U_{\nu}
\ea
$T^{\mu\nu}$ satisfies the conservation equation
$T^{\mu\nu}_{;\mu} = 0$ .
Consider the spatially flat Robertson-Walker metric  
\be
ds^2 = - N dt^2 + e^{2\alpha(t)} dx^i dx^j \delta_{ij}.
\ee
The constraint and equations of motions are
\ba 
& &-6 {\dot\alpha}^2 + 6\dot\alpha \dot\phi + \omega {\dot\phi}^2
  + \rho e^\phi = 0  \no
& &  4\ddot\alpha + 6{\dot\phi}^2 - 2 \ddot\phi +
  {\dot\phi}^2 (\omega + 2 ) - 4 \dot\alpha\dot\phi + p e^{\phi} = 0  \no
& &   6\ddot\alpha + 12 {\dot\alpha}^2 + 2\omega {\dot\phi}^2
   + 6\omega \dot\alpha \dot\phi = 0  
\ea
Solving the energy-momentum conservation 
$\dot\rho + 3\dot\alpha (p + \rho) = 0$ with $p=\gamma\rho$, we get  
$\rho = \rho_{0} e^{-3(1+\gamma)\alpha}$.
Now, we notice that the above equations can be obtained from
\be
S = \int dt\sqrt{N} e^{3\alpha - \phi} [{-6 
\frac{{\dot\alpha}^2}
{N}} + {6 \frac{\dot\alpha\dot\phi}{N}} + {\omega \frac{{\dot\phi}^2}{N}}
 - \rho_0 e^{-3(1+{\gamma})\alpha + \phi} ]
\ee
Introducing a new time $\tau$ by 
\be
d\tau e^{3\alpha - \phi} = dt,  
\ee
the action becomes
\be
 S = \int d\tau [ - \frac{3} {2\sqrt N} (2\alpha^\prime - \phi^\prime)^2 
   + \frac{2\omega + 3}{2\sqrt N} {\phi^\prime}^2 
   - \sqrt{N}\rho_0 e^{3(1-\gamma)\alpha-\phi} ]. 
\ee \\

\section{Some exactly solvable cases}
\subsection{$\gamma=1$ case}
\be
 S = \int d\tau [ - \frac{3} {2} (2\alpha^\prime - \phi^\prime)^2 
   + \frac{2\omega + 3} {2} {\phi^\prime}^2 
   - \rho_0 e^{3(1-\gamma)\alpha-\phi} ]
\ee
We notice that if $\gamma=1$, then we can introduce decoupled variable
$2Y= \phi$ and $X=2\alpha -\phi$ so that we can write the action as
\be
 S = \int d\tau [ -\frac{3}{2}{X^\prime}^2 + 2 (2\omega + 3)
 {Y^\prime}^2 - \rho_{0} e^{-2Y} ] 
\ee
The equations of motions and constraint are
\ba 
X^{\prime\prime} &=& 0  \\ 
Y^{\prime\prime} -  \frac{\rho_0 }{2(2\omega +3)} e^{-2Y} &=& 0  \\
- \frac{3}{2}{X^\prime}^2 + 2(2\omega +3) {Y^\prime}^2 + \rho_{0} e^{-2Y} 
&=& 0
\ea
The solutions of these equations can easiliy be obtained and they are
\ba
X &=& A \tau = 2 \alpha - \phi  \\
Y &=& \log \left ( \sqrt {\frac{2 \rho_{0}}{c(2\omega + 3)}} 
\cosh{\frac{\sqrt{c}}{2} \tau} 
  \right ) = \frac{\phi}{2} 
\ea
From constraint equation the relation between A and c is determined to be
$A= \sqrt {\frac{(2 \omega + 3) c }{3} }$.
By integrating $d\tau e^{3\alpha -\phi} = dt$, we get
\be
t - t_0 = \sqrt{\frac{2\rho_0}{c(2\omega + 3)}}
\left( \f{e^{{\Omega_+}\tau}}{\Omega_+} + \f{e^{{\Omega_-}\tau}}{\Omega_-} 
\right )
\ee
where $\Omega_{\pm} = \sqrt{\f{3(2\o+3)c}{4}} \pm \f{\sqrt c}{2}$.
We define that a function is super-monotonic if 
it is  monotonic and if  the dependent variable cover the entire real line
when the independent variable sweeps the real-line.
We see t is super-monotonic function of $\tau$ if $\omega$ lies between
$-\frac{3}{2}$ and $-\frac{4}{3}$.
These boundary values correspond to $p=0$ and $p=-1$.
The scale factor and dilaton are given as functions of the  $\tau$.
Explicitly, the scale factor and string coupling are
\ba
& &a=e^\alpha = \sqrt{\frac{2\rho_{0}}{c( 2\omega + 3)}} 
          \left (  e^{({\sqrt{\frac{(2\omega + 3)c}{12}} 
            + \frac{\sqrt{c}}{2} ) \tau }}
            + e^{({\sqrt{\frac{(2\omega +3)c}{12}} 
                  - \frac{\sqrt{c}}{2}})\tau} \right ) \\
& &e^\phi =  \frac{2\rho_0}{c(2\omega+3)}   
          \cosh^2(\frac{{\sqrt c}\tau}{2})
\ea
Since $\tau$ and $t$ are super-monotonic, there exists a time $\tau_0$ 
such that $t(\tau_{0})=0$.

Now consider the 0-brane case.
Then $\o= -\f{4}{3}$ and we choose $c=9$ and $6\r_0 = c$ for convenience.
The scale factor and curvature scalar as function of $\tau$ are given by 
\ba
a &=& \left( e^{2\tau} + e^{-\tau} \right ) \no
R &=& 6 (\f{2e^{2\tau} - e^{-\tau}}{(e^{2\tau}
+e^{-\tau})(e^{3\tau} +1)})^2 \no
&+& 6 \f{4e^{2\tau}+e^{-\tau}}{(e^{2\tau}+e^{-\tau})(e^{3\tau}+1)^2} \no
&-& 6\f{3e^{3\tau}(2e^{2\tau}-e^{-\tau})}{(e^{3\tau}+1)^3 
(e^{2\tau}+e^{-\tau})} ,
\ea 
We used $dt = (e^{3\tau} +1) d\tau$.
We notice that as $\tau \rightarrow -\infty$, $t \sim \tau \rightarrow 
-\infty$. In
this limit $a \sim e^{-\tau} \sim e^{-t} \rightarrow \infty$. As 
$\tau \rightarrow \infty$, $t \sim e^{3\tau}$ and
$a \sim e^{2\tau}$.
So, $a \sim t^{\f{2}{3}}$.
At $t \rightarrow -\infty$ the curvature has finite value, scale factor and
coupling go to infinity.
Notice that both scale factor and the curvature is finite in the whole region
so that the singularity is resolved. However the indefinitely growing 
coupling $e^{\phi}$ is not acceptable physically. Therefore the behavior of 
the solution make sense only for the minimum size of the universe, but that is 
what we are looking for.
Long after the big bang (smallest size), many drastic events will happen like 
(non-stringy) inflation, and dilaton will be fixed dynamically, therefore 
the simple model we are considering is no more relevant. Long before the 
big-bang, the physics is unclear at this moment. 
If we take our solution serious, it means that even the 0-branes are not 
a good degree of freedom there are even more bizzare structure of space time 
is waiting. However, one thing is clear:
the universe dominated by the 0-brane is singularity free.
In the figure below, we plot the scale factor, curvature and 
coupling($e^{\p}$) as a function of $\tau$. 
The reason why the cuvature is finite 
and constant is that in the past the scale factor
behaves as a exponential function. 
If $\o$ has other values we can obtain 
the behaviour of curvature such that the maximum of the curvature is located near the minimum of the scale factor.
For example, we set $\o =-1.4$. Integrating 
\be
dt = d\tau 1.29(e^{2.662\tau}+e^{-0.338\tau}) 
\ee
gives 
\be
t = 1.29(\f{e^{2.662\tau}}{2.662} - \f{e^{-0.338\tau}}{0.338}).
\ee 
The scale factor and curvature scalar are
\ba
a &=& 1.29(e^{1.887\tau} + e^{-1.113\tau}) \no
R &=& 6 (\f{1.887e^{1.887\tau}-1.113e^{-1.113\tau}}{1.29(e^{1.887\tau}
+e^{-1.113\tau})(e^{2.662\tau} + e^{-0.338\tau})})^2 \no 
&+& 6 \f{1.887^2 e^{1.887\tau} + 1.113^2 
e^{-1.113\tau}}{1.29^2(e^{1.887\tau}
+e^{-1.113\tau})(e^{2.662\tau}+ e^{-0.338\tau})^2} \no
&-& 6 
\f{(2.662e^{2.662\tau}-0.338e^{-0.338\tau})(1.887e^{1.887\tau}-1.113e^{-1.113\tau})}
{(1.29^2(e^{1.887\tau}+e^{-1.113\tau})) (e^{2.662\tau}+e^{-0.338\tau})^3} 
\ea

\begin{figure}
\begin{center}
$${\epsfxsize=10.0 truecm \epsfbox{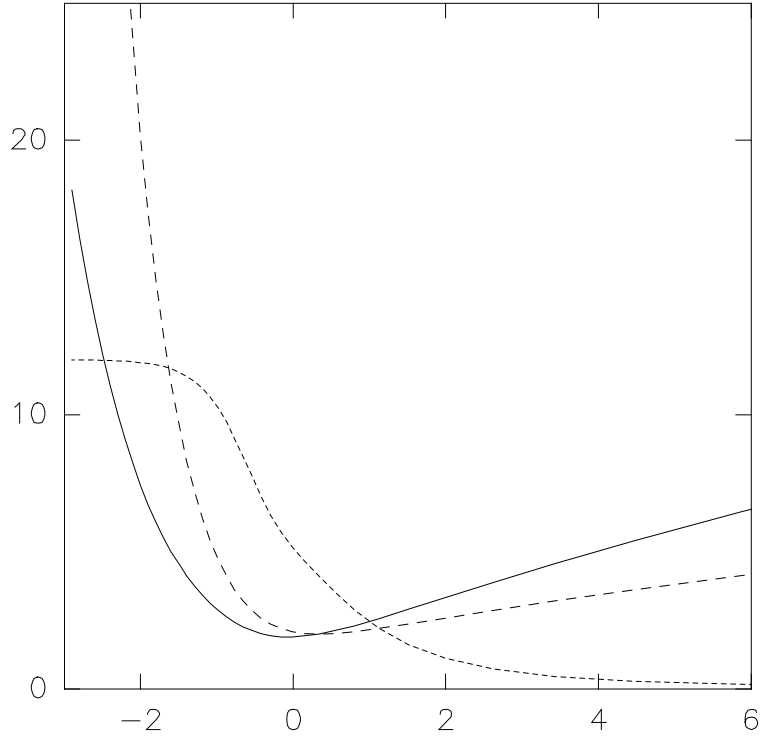}}$$
\caption{
 Behaviour of $a=e^{\alpha}$, $R$ and $e^{\phi}$ in string frame.
I have set $\omega = -\frac{4}{3}$, $c=9$ and $6\rho_{0} =c$.
{\it{Solid line:scale factor$(a)$, Dotted line:curvature$(R)$, Dashed 
line:coupling$(e^{\phi})$}}
}
\end{center}
\end{figure}

In figure below we see the curvature scalar when $\o=-1.4$.

\begin{figure}
\begin{center}
$${\epsfxsize=10.0 truecm \epsfbox{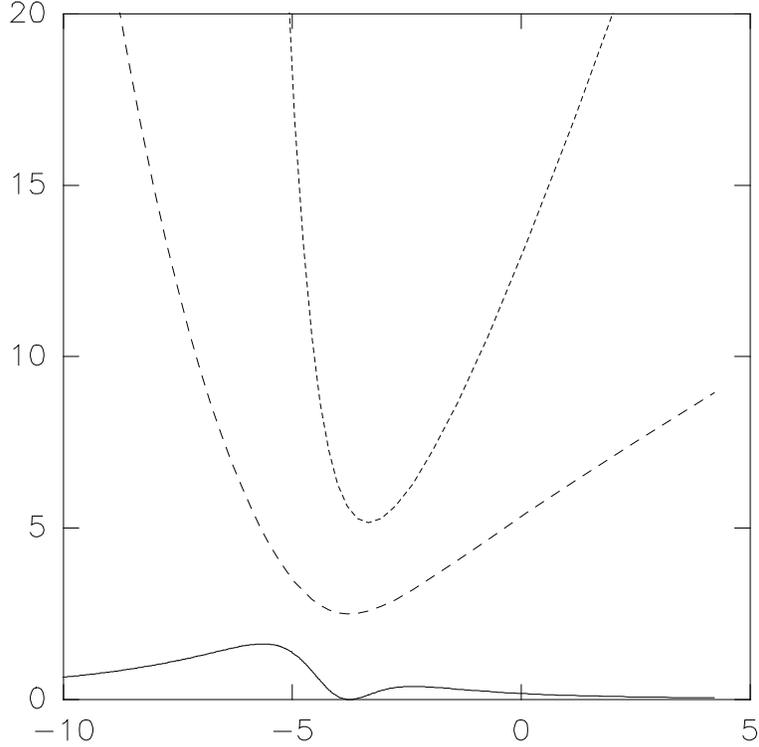}}$$
\caption{
 The curvature scalar when $\o =-1.4$. {\it{Solid
line:Curvature, Dotted line:coupling, Dashed line:scale factor}}.
}
\end{center}
\end{figure}

For string, $\o =-1$,  $t$ and $\tau$ relation is
\be
t-t_0 = \sqrt{\f{\r_0}{2c}} \left(\f{e^{{\sqrt 
\Omega_{+0}}\tau}}{\Omega_{+0}}
+\f{e^{{\sqrt \Omega_{-0}}\tau}}{\Omega_{-0}} \right ) 
\ee
where, $\Omega_{\pm 0} = \f{\sqrt 3c}{2} \pm \f{\sqrt c}{2}$.
So, the relation of $t$ and $\tau$ is not strongly monotonic. This is the 
origin 
of singularity. As $\tau$ runs $-\infty$ to $\infty$, $t$ runs 
$0$ to $\infty$. 
During this time the scale factor runs $0$ to $\infty$. More explicitly
\be
a=e^{\a} = \sqrt{\f{\r_0}{2c}} \left( e^{({\sqrt \f{c}{12}} 
+\f{\sqrt{c}}{2})\tau}
+e^{({\sqrt{\f{c}{12}} - \f{\sqrt c}{2}})\tau} \right ) 
\ee
Since  the scale factor goes to zero in a finite conformal time $t$, we find
singularity in a finite time. It is a true curvature singularity. \\ 

Since we are talking about near the big-bang, one might
think that the universe is still
in the uncompactified state.
So, in this paragraph we work in 10-dim or any dimension D-dim. 
\be
ds_{10}^2 = -dt^2 + e^{2\a} dx_i dx^i (i = 1,2,3, \cdots, 9)
\ee
then the curvature is $R = 18{\ddot\a} + 90 {\dot\a}^2$.
the action in 10-dim.
\be
S=\int dt e^{9\a- \p} \left ( -72 {\dot\a}^2 + 18 \dot\a \dot\p
+ \o {\dot\p}^2 - \r_0 e^{-9(1+\g)\a -\p} \right)
\ee
Introducing $d\tau e^{9\a-\p} = dt$.
\ba
S &=& d\tau \left( -72 {\a^\prime}^2  + 18\a^\prime \p^\prime 
+\o{\p^\prime}^2
-\r_0 e^{9(1-\g)\a -\p}\right) \no
&=& \int d\tau \left(-72(\a^\prime-\f{\p^\prime}{8})^2 
+\f{8\o+9}{8}{\p^\prime}^2 -\r_0e^{-\p} \right)
\ea
where we set $\g=1$.
introducing new variables $\a^\prime - \f{\p^\prime}{8} = X$ and
$\p = 2Y$,
\be
S= \int d\tau \left(-72{X^\prime}^2 +\f{8\o +9}{2}{Y^\prime}^2 
-\r_0e^{-2Y} \right).
\ee
The constraint and equations of motion are 
\ba
-72{X^\prime}^2 + \f{8\o+9}{2}{Y^\prime}^2 - \r_0e^{-2Y} = 0 \no
Y^{\prime\prime} - \f{2\r_0}{8\o+9} e^{-2Y} = 0 \no
X^{\prime\prime} = 0.
\ea
The solutions are
\ba
Y = \log [ \sqrt{\f{8\r_0}{(8\o+9)c}} \cosh{\f{\sqrt c}{2}\tau} \no
X = A \tau
\ea
where $A = \sqrt{\f{(8\o+9)c}{576}}$.
If $\o$ satisfies $-\f{9}{8} < \o \le -\f{10}{9}$, then the relation
$t$ and $\tau$ is strongly monotonic.
Since $9\a -\p$ is given by $9A\tau +\f{\p}{8}$, integration
$d\tau e^{9\a-\p} = dt$ gives
\be
\int dt = \int d\tau (constant) e^{\sqrt{\f{81(8\o+9)c}{576}}\tau}
(\cosh{\f{\sqrt c}{2}\tau})^{\f{1}{4}} 
\ee
If $\o = -\f{10}{9}$,
\be
t \sim \int d\tau e^{\f{c}{8} \tau}(\cosh{\f{\sqrt c}{2}\tau})^{\f{1}{4}}
\ee
So, as $t \rightarrow -\infty$, $t \sim \tau$. That is $t$ is
strongly monotonic function of $\tau$.
This result can be extended to arbtrary dimension $D$, the action is
\be
S= \int dt e^{(D-1)\a -\p} \{ (D-1-(D-1)^2){\dot\a}^2 +2(D-1)\dot\a\dot\p
+\o{\dot\p}^2 -\r_0e^{-(D-1)(1+\g)\a+\p} \} 
\ee
For $\g=1$,
\be
S= \int d\tau \left [ -\f{(D-1)}{(D-2)}((D-2)\a^\prime - \p^\prime)^2
+\f{(D-2)\o +(D-1)}{D-2}{\p^\prime}^2 -\r_0 e^{-\p} \right ]
\ee
If $\o$ satisfies $-\f{D-1}{D-2} < \o \le -\f{D}{D-1}$, then
we have nonsingualr solution. \\

\noindent
\subsection{$\gamma = \frac{2}{3}$ case}  
One can adopt similar procedure as above to get 
\be
t-t_{0} = \int d\tau \left( 
(\f{\vl{\o}\vl\r_{0}}{3c(2\o+3)})^{\f{12}{\vl{\o}\vl}}
e^{\sqrt{\frac{12(2\omega+3)c}{{\vline\omega\vline}^2} \tau}} 
(\cosh{\frac{\sqrt c}{2} \tau})^{\frac{12}{\vline\omega\vline} - 6} \right)
\ee
If $\omega$ lies between $-\frac{3}{2} < \omega \le -\frac{4}{3}$, 
then the $t$ and $\tau$ relation is monotonic.
For example, if we take $\o$ as $-\f{4}{3}$ then 
\be
t-t_0 =\int d\tau ( (\f{4\r_0}{3c})^{\f{5}{2}} e^{\f{3\sqrt c}{2}\tau}
 (e^{\f{\sqrt c}{2}\tau} + e^{-\f{\sqrt c}{2}\tau})^3 ) ,
\ee
and scale factor
\be
a=e^{\a} =(\f{4\r_0}{3c})^{\f{5}{2}}e^{\f{3\sqrt c}{4}\tau}
      (e^{\f{\sqrt c}{2}\tau} + e^{-\f{\sqrt c}{2}\tau})^{\f{5}{2}}
\ee
As $\tau \rightarrow \infty$ or $-\infty$, $t \sim e^{{\sqrt c}\tau} 
\rightarrow \infty$ or $t \sim -e^{{-\sqrt c}\tau} \rightarrow -\infty$ 
and  
scale factor as $t \rightarrow \infty$, $a \sim e^{2{\sqrt c}\tau} 
\rightarrow \infty$
or as $t \rightarrow -\infty$, $a \sim e^{-\f{\sqrt c}{2} \tau} \sim 
(-t)^{\f{1}{2}}$.
From the fact that $t$ and $\tau$ are monotonic, it is right there is 
a finite time $\tau_0$ where $t=0$.  
With this fact we see that the scale factor does not touch zero 
in a finite cosmic time, so it is singularity free.
For the coupling $e^\p$,
\be
e^\p = \sqrt{\f{4\r_0}{3c}} e^{\f{3\sqrt c}{4}\tau} (e^{\f{\sqrt c}{2}\tau}
+ e^{-\f{\sqrt c}{2}\tau} )^{\f{9}{2}}
\ee
As $\tau \rightarrow \infty$ or $-\infty$, $t \sim \infty$ or $-\infty$
then $e^\p \sim \infty$. The behaviour of diverging coupling 
does not match if we compare to our today's universe. 
However, we are looking near the big-bang not late time universe. 
The non-singularity could not be obtained if we considered $\o=-1$ 
because when $\o = -1$, $t$ behaves $t \sim e^{(2\sqrt {3c} - 3\sqrt c) 
\tau}
\rightarrow 0$ for $\tau \rightarrow -\infty$ and $t \sim e^{(2\sqrt {3c}
+3\sqrt c)\tau} \rightarrow \infty$ for $\tau \rightarrow \infty$.
As we have seen before, the super-monotonity determines whether 
the solution is singular or not. \\ 

\noindent\subsection{$\g = \f{1}{3}$} 
In this case,
\be
t-t_{0} = \int d\tau \sqrt {\f{2\rho_{0}}{3c}} e^{\sqrt{\f{3c}{4(2\o + 3)}}\tau}
        (\sinh {\f{\sqrt c}{2} \tau})^{-3}
\ee
The scale factor and coupling is given as function of $\tau$
\ba
a= e^{\a} = \sqrt{\f{3c}{2\r_0}}e^{\sqrt{\f{3c}{4(2\o+3)}}\tau}
  (\sinh {\f{\sqrt c}{2}\tau})^{-1} \no
e^{\p} = e^{\sqrt{\f{3c}{2\o+3}}\tau}
\ea

We can easily see that $t$ and $\tau$ relation is not strongly monotonic for
any $\o$. 
Also, both the scale factor and the conformal time diverge at $\tau=0$.
As $\tau \rightarrow \pm \infty$, $t \rightarrow 0$ or $t \rightarrow \infty$.
As before this monotonity says the scale factor has a singularity because 
conformal time cannot be defined from $-\infty$ to $\infty$. \\

{\bf 4. Solution coupled to metter $\g = -1$ : vacuum energy density} \\
In this case, 
the relation $\int dt=\int d\tau e^{3\a -\p}$ gives
\be
\int dt = \int d\tau \sqrt{\f{4\rho_{0}}
{-4c\mid \o \mid +\f{c(12\o+6)^2}{\vl{36\o+30}\vl}}}e^{-3A\tau} 
(\cosh{\f{\sqrt c}{2}\tau})^{-2+3\f{\vl{12\o+6}\vl}{\vl{36\o+30}\vl}}
\ee
From this relation, we find there are no $\o$'s to make super-monotonity.
To get monotinity from like above integration, we must have $\o$ satisfying
$-2+\f{\vl{12\o+6}\vl}{\vl{36\o+30}\vl} \ge -3A\tau$ and
$-2+ \f{\vl{12\o+6}\vl}{\vl{36\o+30}\vl} \ge 0$. There are no such $\o$'s.
As $\tau$ goes to $\pm \infty$, $t$ goes to zero. 
So, this result is also contain curvature singularity.

We summarize what we have obtained by making a table.
\begin{itemize}
\item {\bf {Summary}}
\end{itemize}
\vskip 0.5cm
\begin{tabular}{|c||c|c|c|c|} \hline
 & $\gamma=1$ & $\gamma=\frac{2}{3}$ & $\gamma=\frac{1}{3}$ & $\gamma= -1$ 
\\ \hline
$t(\tau = -\infty)$ & $-\infty$ & $-\infty$ & $0$, 
singular at $\tau =0$ & $0$ \\ \hline
$a(\tau = -\infty)$ & $ e^{-t}$ & $(-t)^{\f{1}{2}}$ & $ 0 $, singular at $\tau = 0$ & $0$ \\ \hline
$a(\tau = \infty)$ &$t^{\frac{2}{3}}$ & $t^{\frac{2}{3}}$ &$ e^{3 t}$ & $t^{-1.77} $ \\ \hline
R(curvature) & non-singular & non-singular & 
singular &singular \\ \hline
\end{tabular}

\section{ Einstein frame}
\noindent
\subsection{\bf Effective action in Einstein frame:} 
For completeness we study the problem in a Einstein frame and 
analyze a few examples and compare with corresponding solutions 
of the Brans-Dicke frame.
To obtain Einstein frame action, we have to do conformal rescaling 
$g_{E\mu\nu} = e^{-\p} g_{\mu\nu}$.
Then after transformation the action becomes
\be
S_{E} = \int d^4x \sqrt{-g_{E}} [ R_{E} -\f{2\o+ 3}{2} (\partial\p)^2 ] 
+S_m
\ee
By varying $g_{E\mu\nu}$ and $\p$ we obtain equations of motion.
\ba
R_{E\mu\nu} - \f{1}{2} g_{E\mu\nu} R_{E} + \f{(2\o+3)}{4} (\partial\p)^2 
- \f{2\o+3}{2}\partial_{\mu}\p\partial_{\nu}\p = T_{\mu\nu} \no
\partial_{\mu}(\sqrt {-g_{E}} \partial^{\mu}\p) = 0
\ea 
where,
\be
T_{\mu\nu\p} = -\f{2\o+3}{4}g_{E\mu\nu}(\partial\p)^2 +\f{2\o+3}{2}
\partial_\mu\p \partial_\nu\p. 
\ee
Notice that in Einstein frame we clearly see why the condition 
$2\o + 3 > 0$ should be satisfied. This condition reproduces 
the Weak Energy Condition(WEC) : $\r_E > 0$, $\r_E + p_{Ei} > 0$. 
The sign of $2\o +3$ must be chosen to satisfy WEC.

If we take spatially flat metric $ds_E^2 = -dt_E^2 
+ e^{2\a_E} dx_i dx^j$ so, $R_{E00} =-3\f{\ddot a_E}{a_E}$ and
$R_E= 6(\f{\ddot a_E}{a_E} + (\f{\ddot a_E}{a_E})^2)$, 
then we get the equations of motion.
\ba
3{\dot\a_{E}}^2 -\f{2\o+3}{4}{\dot\p}^2 = \r_{E} \no
-2{\ddot\a_{E}} - 3{\dot\a_{E}}^2 - \f{2\o+3}{4}{\dot\p}^2 = p_{E} \no
{\ddot\p} + 3{\dot\a_{E}}{\dot\p} = 0
\ea
We use $\r_{E} = \r_{E0} e^{-3(1+\g)\a_{E}}$ of the solution 
$\dot\r_{E} + 3{\dot\a_{E}}(\r_{E} + p_{E}) = 0$ derived from
continuity equation $T_{;\mu}^{\mu\nu} = 0$. 
Furthermore, we see again that the equations can be derived from
\be
S_E = \int dt_E e^{3\a_E} (-6{\dot\a_E}^2 + \f{2\o+3}{2}{\dot\p}^2
    -2\r_{E0} e^{-3(1+\g)\a_E} )
\ee
Like string frame, using $d\tau_E  e^{3\a_E} = dt_E$ then
\be 
S_E = \int d\tau_E (-6{\a_E^\prime}^2 +\f{2\o+3}{2}{\p^\prime}^2 -2\r_{E0} 
e^{3(1-\g)\a_{E}} )
\ee  
When $\o=-1$, the action becomes familiar to us
\be
S_E = \int d\tau_E (-6{\a_E^\prime}^2 + \f{1}{2}{\p^\prime}^2 -\r_{E0}
e^{3(1-\g)\a_E} )
\ee
If we find solution in string frame then we can transform to Einstein frame
by some calculation and vice versa.
Since $g_{E\mu\nu} =e^{-\p} g_{\mu\nu}$, metric is given by $ds_E^2 =
ds^2 e^{-\p}$. By using these
relations we can transform solutions in one frame to the other. 
However, we must be careful when matter is coupled. Solutions which is
obtained by directly solution transformation would be different from those 
of obtained by transformation. \\

In the following, we will consider solutions in Einstein frame. The 
solution
in this frame can be obtained as explained above by rescaling string frame
metric. This makes it possible to transform solutions in string frame to
Einstein frame and vice versa. Since we found solutions in string frame,
let's transform into Einstein frame. In Einstein frame the metric is   
\ba
ds_E^2 &=& e^{-\p}(-dt^2 + e^{2\a}dx_i dx^i) \no
  &=&-dt_E^2 + e^{2\a_E}dx_idx^i
\ea
From the above relation we find $\a_E = \a - \f{\p}{2}$ and
$dt_E = e^{-\f{\p}{2}}dt$. If we found solution in one frame,
then by these relations we can obtain solution in the other frame. 

\subsection{$\g=1$ : massless scalar fields} 
We will compare transformed solution from string frame to the one
that obtained transformed action. First, consider transformed solution 
from string frame.
Since $dt_E = e^{-\f{\p}{2}} dt = e^{3(\a-\f{\p}{2})}d\tau = d\tau 
e^{\f{3A}{2}\tau}$, integration gives
\be
t_E = \f{2}{3A} e^{\f{3A\tau}{2}} 
\ee
Then, the transformed scale factor into Einstein frame is 
\be
a_E = e^{\a_E} = e^{\f{A}{2}\tau} =(\f{3A}{2}t_E)^{\f{1}{3}}
\ee
We obtained singular solution transformed from
that of non-singular. This singularity 
originates from the divergence of the dilaton at the $-\infty$ in $\tau$.
    
Now, let's start from action in Einstein frame. For $\g=1$ we can write the 
action
\be
S_E = \int d\tau_E (-6 {\a_E^\prime}^2 + \f{2\o+3}{2}{\p^\prime}^2 
-2\r_{E0} )
\ee
The constraint and equations of motion are
\ba
-6{\a_E^\prime}^2 + \f{2\o+3}{2}{\p^\prime}^2 + 2\r_{E0} = 0 \no
\a_E^{\prime\prime} = 0 \no
\p^{\prime\prime} = 0
\ea
The solutions are $\a_E=A\tau_E$ and $\p = B\tau_E$ where $A$ and $B$
are related by $-6A^2 +\f{2\o+3}{2}B^2 + 2\r_{E0} = 0$.
In this case we also found singular solution similar to what we obtained by
transformation. Integration $dt_E = d\tau_E e^{3\a_E}$ gives
$t_E \sim e^{3A\tau_E}$ so cosmic time $t_E$ spans at least $0$ to 
$\infty$.
Scale factor $e^{\a_E}(\sim t_E^{\f{1}{3}})$ goes to zero as 
$t_E$ goes to zero.
So, we find there does not exist non-singular solution. It is certain
that the scale factor that touches zero in a finite time has a curvature
singularity. \\

\noindent\subsection{$\g=\f{2}{3}$} 

From solution of string frame,
\ba
\int dt_E &=& \int e^{-\f{\p}{2}} dt = \int d\tau e^{3(\a - \f{\p}{2})} \no
    &=& \int d\tau e^{3\a_E} = (cosnt) \int d\tau e^{\f{9{\sqrt c}}{8}\tau}
(e^{\f{\sqrt c}{2}\tau} + e^{-\f{\sqrt c}{2}\tau}) 
\ea
As $\tau \rightarrow \pm \infty$, $t_E$ runs $0$ to $\infty$. 
It is singular.
For scale factor 
\be
a_E = e^{\a_E} = (const) e^{\f{3\sqrt c}{8}\tau}
(e^{\f{\sqrt c}{2}\tau} + e^{-\f{\sqrt c}{2}\tau})
\ee
We see, as expected, $a_E$ runs from $0$ to $\infty$ as conformal time $t_E$
runs from $0$ to $\infty$. It is truly singular. We also obtained sigular 
solution like former case by transforming that of non-singular in string frame.

For $\gamma=-1$,   
we start with transformed action, 
\be
S_E = \int d\tau_E \left( -6{\a_E^\prime}^2 + \f{2\o +3}{2}{\p^\prime}^2 
      - 2\r_{E0}e^{6\a_E} \right) = 0
\ee
In order to satisfy constraint equation, we take the solution 
\be
-3\a_E = \log [\sqrt {\f{12\r_{E0}}{c}}\sinh{\f{\sqrt c}{2} \tau_E} ]. 
\ee
Again, the relation $t_E$ and $\tau_E$ 
is not monotonic. At $\tau_E = 0$ transformed action 
also gives singular solution.

From above cases, we expect there are no singularity free solution 
in the Einstein metric, as expected.

\section {Discussion and conclusion}

In this paper we solve for some fixed values of $\g$ in the Brans-Dicke 
action.
We assumed that the universe is dominated by 
one kind of p-brane of perfect fluid type.
We found an analytic solution  which is singularity free 
for some $\o$ and $\g$. 0 branes and instantons
are so important near the singular region by which our universe can avoid
singularity.
We conjecture that for matters $\g$ in the region
$-1 \le \g \le \f{1}{3}$, we cannot avoid singularity. For 
matters in the region $\f {1}{3} < \g < \f{2}{3}$ we
still need more study. 

We out-line remainging problems for later research. 
 Firstly, once p is fixed, presumably both $\omega$ and 
  $\gamma$ should be fixed. 
Without knowing the parameter for the fixed p, 
we have to classify ossible cosmology according to the parameter of 
the system. This will be reported elsewhere\cite{park}.
 Secondly and more importantly, we do not have a rigorous basis for the cosmology of string solitons.
If what we took as basis goes wrong, then what we have done is just Brans-Dicke
cosmology in the presence of some perfect fluid type matter.
We wish that more study of the effect of the solitons in the string 
cosmology be done in the future.

\vskip 1cm
\noindent{\bf \Large Acknowledgement}

\noindent This work has been supported by the resesarch grant KOSEF 971-0201-001-2.

\end{document}